\documentclass[11pt,fleqn]{article}  
\usepackage{psfig}
\title{Features of the Acoustic Mechanism of Core-Collapse
Supernova Explosions : Revisited using the Kinetic Approach }
\author{Zhu Guang Hua} 
\date{Department of Physics, Wulumuqi Toudiban (Transfer Centre), Xinjiang 830000, China}
\begin{document}      
\maketitle                 
\begin{abstract}           
The discrete kinetic model is used to study the propagation of sound
waves in system of hard-disk-like rotating stars (or vortex gases).
The anomalous (negative) attenuation or amplification which is
possibly due to the binary collision of a dilute-enough rotating
disk (or vortex-gas) system (each with opposite-sign rotating
direction or angular momenta but the total (net) angular momenta or
vorticity is zero) or microreversibility might arise from the
implicit balance of the angular momentum during encounter and give
clues to the understanding of possible acceleration of cosmic rays
passing through this kind of channel and direct or inverse
vortex-gas cascades in two-dimensional turbulence of astrophysical
problems.

\vspace{2mm} \noindent Keywords : physics of the early universe,
classical tests of cosmology, neutrino detectors, ultra high energy
cosmic rays, core-collapse supernovas

\end{abstract}
\doublerulesep=6.5mm    
\baselineskip=6.5mm  
\section{Introduction}
\bibliographystyle{plain}
There are many proposed acoustic probes in astrophysical
applications, say, studying baryon acoustic oscillations or
microwave background fluctuations (due to couplings between
different cosmic background particles) [1]. Recently Burrows {\it et
al.} [2] proposed a new mechanism for core-collapse supernova
explosions : the acoustic power generation in the thin core as the
driver and the propagation into the mantle of strong sound waves.
Acoustic power is, potentially, an efficient means to transport
energy and momentum into the outer mantle to drive the supernova
explosion. Unlike neutrinos, sound is almost 100\% absorbed in the
matter, though some of the acoustic energy is reradiated by
neutrinos. As sound pulses propagate outward down the density
gradient they steepen into multiple shock waves that catch up to one
another and merge; a shock wave is almost a perfect black-body
absorber of sound. If sufficient sound is generated in the core, it
would be a natural vehicle for the gravitational energy of infall to
be transferred to the outer mantle and could be the key missing
ingredient in the core-collapse explosion mechanism.
\newline
Using the 2D radiation/hydrodynamic code VULCAN/2D Borrows {\it et
al.} [2] meanwhile have discovered that turbulence and anisotropic
accretion in the inner 40-100 kilometers can excite and maintain
vigorous core g-mode oscillations which decay by the radiation of
sound. The inner core acts as a transducer for the conversion of
accretion gravitational energy into acoustic power. As they used
two-dimensional treatment, thus they are adopting gases of disks
instead of gases of spheres [2]. In fact, there are other approaches
using the kinetic or Boltzmann equations to study neutrino physics
[3-5].
\newline Quite recently the search for ultra high energy (UHE,
10$^{18-20}$ eV) neutrinos is strongly motivated by the observation
of cosmic rays. Since the first comprehensive observations by Auger
in 1938, it has been observed that by extending the coverage of
detectors, increasingly higher energy events have been detected.
Presently, the spectrum of these outer-space particles is known to
extend up to 10$^{20}$ eV, and maybe even higher. In particular,
there is an on-going discussion on the existence, or not, of the GZK
[6] cut-off on the cosmic rays spectrum. Because neutrinos interact
so weakly, their observation at UHE would bring insight on the
origin and nature of these UHE cosmic rays. If the GZK cut-off is
relevant, it could be studied from the observation of neutrinos
which will be produced as secondaries from pion disintegrations.
This would signal the existence of extremely high energies cosmic
particles (EHE, 10$^{20+}$ eV). Furthermore, neutrinos are the only
known candidates that would allow astrophysical observation at these
extreme energies, above the GZK cut-off [8]. There already are many
intensive researches reported till now [9].
Meanwhile, as reported in [10], there is a significant
transformation of kinematic energy into compressible sound energy
when the (superfluid Helium) vortex tangle evolves. This might imply
that the production of sound is the fundamental process of
dissipation of kinetic energy at absolute zero (T=0) [10]. Thus,
second sound is an important tool in the research related to quantum
turbulence. They can also be applied to the relevant problems in
astrophysics or cosmology [11] becaus all these vortex-like studies
are closely linked to system of rotating hard-disk stars. Thus,
better understanding of the formation and/or decay of vorticity in
the BEC and the interaction of multiple vortices in dilute Bose
gases (or stirred BEC) subjected to thermal noises currently raise
many challenges [12]. Either collision or shearing of vortices
during encounters will generate thermodynamic noises or sound
pulses. Eventually the unstable or turbulent state in BEC occurs.
\newline In fact, a vortex could be treated as a combination of
concentrated vorticity (core) with its surrounding irrotational
fluid (flow) [13]. Thus, they could be, in certain sense, similar to
the hard-sphere particles when the elastic scattering or collision
for a system of them being in consideration [14]. Some researchers
may argue that, the basic physics of vortices is, according to a
brilliant formulation of Onsager [15], that configuration space and
phase space coincide. So, in 2D, the x-coordinate and the
y-coordinate of a vortex are conjugate variables! This has important
consequences for identifying the momentum of a vortex, i.e. a vortex
never behaves kinematically like a Newtonian point particle (hard
sphere in 3D or hard disc in 2D). System of vortices [16-17], dilute
enough and during binary encounters, however, as the chaotic
behavior [18-19] resembling that of Newtonian particles  upon
collisions [13-14], could be treated dynamically by hard-sphere or
hard-disk collisions with careful selection of the impact parameter.
\newline First sound is well known as an ordinary sound wave and is
related to a wave of pressure difference (or a kind of fluctuation
in the total density) in helium II. Second sound is unique to helium
II and represents a kind of fluctuation in temperature. Both share
similarities to the propagation of sound mode (first mode) and
diffusion mode (second mode; entropy wave) in dilute monatomic gases
or fluids which were already well captured by continuous and/or
discrete kinetic models [20-22] using hard-sphere- or hard-disk-gas
assumptions.
\newline Recent Jackiw gave hydrodynamic profiles and constants of
motion from d-branes and discussed supersymmetric fluid mechanics
[23]. It will be interesting to further investigate other
hydrodynamic properties, like sound propagation in dilute gases with
similar hard-disk like collisions  for nonrelativistic cases. As a
preliminary attempt, considering the presumed analogy between the
scattering of dilute rotating (vortex) gases [24-27] (e.g., the
duality or equivalence of Bose-gas particles [24-25] and (rotating)
vortices [26-27]) and dilute hard-disk gases, in this paper, we plan
to investigate the dispersion relations of sound propagation in
hard-disk-like rotating stars by the orientation-free discrete
kinetic model which has been verified in Refs. [21-22,28]. This
presentation will give some clues to the understanding of the
dissipation mechanism of interactions of multiple rotating
hard-disk-like stars (or vortices) and other similar problems in
astrophysics or cosmology [1-9,11] (acoustic probes for
two-dimensional turbulence included). We note that complicated
boundary conditions are avoided by most of previous workers who did
initial value problems in an unbounded domain rather than the
semi-infinite problem to which measurements refer because of the
essential difficulties.
\newline In the discrete kinetic approach, the main idea is to
consider that the particle velocities belong to a given finite set
of velocity vectors. Only the velocity space is discretized, the
space and time variables are continuous. The discrete kinetic models
[21-22,28] thus come (please see the detailed references therein).
For a lattice gas, the space and time variables are also
discretized. By using the discrete kinetic approaches, the velocity
of propagation of sound wave can be classically determined by
looking for the properties of the solution of the conservation
equation referred to the Maxwellian state. Theoretical attempts link
quantum-mechanic to Boltzmann approach have been well developed
(see, e.g., [22]) which provide us more applications for the present
approach. \newline In this presentation, we shall introduce our theoretical
approach in the Section 2. The numerical results and discussions will be
put in the final Section. We shall firstly present those fixed-orientation
results and then other remaining free-orientation results which might be useful
to interpret those claims in [2]. 
\section{Theoretical Formulations} We make the following assumptions before we
investigate the general equations of our model : \newline (1)
Consider a gas of identical particles of unit mass and a shape of
a disk of diameter $d$, then each particle $i$, $i=1,\cdots,N$, is
characterized by the position of its center $q_i$ and its velocity
$v_i$. We also have the geometric limitations : $|q_i -q_j|$ $\ge
d$, $i\not =j$.
\newline (2) Each particle moves in the plane with velocity
belonging to a discrete set ${\cal V}$ of 4 velocities with only one
speed in the plane. The velocity modulus $c$ is a reference speed
depending on the reference frame and specific distribution of
particles. $c$ is normally linked to the internal energy of the
molecules in thermodynamic equilibrium (please see Fig. 2).
\newline (3) The collisional mechanism is that of rigid spheres,
that is, the particles scatter elastically and they change their
phase states instantaneously, preserving momentum. Only binary
collisions are considered, since a multiple collision here is a
negligible events.
\newline The collisions between two particles (say $i$ and $j$)
take place when they are located at $q_i$ and $q_j=q_i-d {\bf n}$,
where ${\bf n}$ is the unit vector joining their centers. After
collisions the particles scatter, preserving momentum, in the
directions allowed by the discrete set ${\cal V}$. In other words,
particles change according to
\begin{displaymath}
 (q_i, v_i) \rightarrow (q_i, v^*_i), \hspace*{10mm}
 (q_j, v_j) \rightarrow (q_j, v^*_j) .
\end{displaymath}
The collision is uniquely determined if the incoming velocity and
the impact angle $\psi$, $\psi \in$ [$-\pi/2$,$\pi/2$], are known,
which is defined as the angle between $v_i$ and ${\bf n}$ or ${\bf
n}(\psi)$ = ($\cos$ [$\psi+(k-1)\pi/2$], $\sin$
[$\psi+(k-1)\pi/2$]), $k=1, \cdots, 4$.  \newline From the
selected velocities we have two classes of encounters, i.e.
$\langle v_i, v_j \rangle$ = $0$  and $\langle v_i, v_j \rangle$ =
$-c^2$, respectively. \newline (a). In the first class momentum
conservation implies only : encounters at $\pi/2$ with exchange of
velocities
\begin{displaymath}
 v_i =v^k \rightarrow v_i^* =v^{k+1}, \hspace*{10mm} v_j =v^{k+1}
 \rightarrow v_j^* =v^k, \hspace*{4mm} k=1,\cdots, 4,
\end{displaymath}
in the case $\psi \in [ - \pi/2, 0]$, and
\begin{displaymath}
 v_i =v^k \rightarrow v_i^* =v^{k+3}, \hspace*{10mm} v_j =v^{k+3}
 \rightarrow v_j^* =v^k,
\end{displaymath}
in the case  $\psi \in$ $[0, \pi/2]$.  \newline (b) Similarly,
$\langle v_i, v_j \rangle$ = $-c^2$; \newline (i) Head-on
encounters with impact angle $\psi=0$ such that
\begin{displaymath}
 v_i =v^k \rightarrow v_i^* =v^{k+2}, \hspace*{10mm} v_j =v^{k+2}
 \rightarrow v_j^* =v^k, \hspace*{4mm} k=1,\cdots, 4,
\end{displaymath}
(ii) Head-on encounters with impact angle $\psi \not=0$ such that
\begin{displaymath}
 v_i =v^k \rightarrow v_i^* =v^{k+1}, \hspace*{10mm} v_j =v^{k+2}
 \rightarrow v_j^* =v^{k+3}, \hspace*{6mm} \mbox{if $\psi \in$ $[-\pi/2,
 0]$} ,
\end{displaymath}
\begin{displaymath}
 v_i =v^k \rightarrow v_i^* =v^{k+3}, \hspace*{10mm} v_j =v^{k+2}
 \rightarrow v_j^* =v^{k+1}, \hspace*{6mm} \mbox{if $\psi \in$ $[0,
 \pi/2]$}.
\end{displaymath}
For grazing collisions, that is $\langle {\bf n}, v_i \rangle$=
$\langle {\bf n}, v_j \rangle$ = $0$ , we put $v^*_i =v_i$, $v_j^*
=v_j$. Schematic presentation is illustrated in Figs. 1 and 2.
\newline \setlength{\unitlength}{1.0mm}
\begin{picture}(120,65)(-30,-30)
\thinlines \put(7,-3){\circle{11.0}} \put(12,7){\circle{11.0}}
\thicklines \put(7,-3){\vector(1,-1){15}}
\put(7,-3){\vector(1,2){16}} \put(7,-3){\vector(2,1){15}}
\put(-7,12){\makebox(0,0)[bl]{\large {\bf $v_1$}}}
\put(-9,0){\makebox(0,0)[bl]{\large {\bf $v^*_1$}}}
\put(12,7){\vector(-3,2){15}} \put(12,7){\vector(-2,-1){15}}
\put(24,1){\makebox(0,0)[bl]{\large {\bf $v$}}}
\put(12,-15){\makebox(0,0)[bl]{\large {\bf $v^*$}}}
\thinlines \put(0,-3){\line(4,0){30}} \put(12,7){\line(2,0){15}}
\put(17,22){\makebox(0,0)[bl]{\large {\bf $n$}}}
\put(-30,-32){\makebox(0,0)[bl]{\small {Fig. 1 \hspace*{2mm}
Schematic (plot) of a binary collision.}}}
\end{picture}      
\begin{picture}(120,65)(80,-30)
\thinlines \put(77,-3){\circle{11.0}} \put(82,7){\circle{11.0}}
\thicklines \put(77,-3){\vector(1,-1){15}}
\put(77,-3){\vector(1,2){16}} \put(77,-3){\vector(2,1){15}}
\put(63,12){\makebox(0,0)[bl]{\large {\bf $v^*_1$}}}
\put(61,0){\makebox(0,0)[bl]{\large {\bf $v_1$}}}
\put(82,7){\vector(-3,2){15}} \put(82,7){\vector(-2,-1){15}}
\put(94,1){\makebox(0,0)[bl]{\large {\bf $v$}}}
\put(82,-15){\makebox(0,0)[bl]{\large {\bf $v^*$}}}
\put(87,22){\makebox(0,0)[bl]{\large {\bf $n$}}}
\put(60,-32){\makebox(0,0)[bl]{\small {Fig. 2\hspace*{2mm} A {\it
head-on} collision.}}}
\end{picture}
\vspace{4mm}

\noindent The verification of our approach with the previous
available approaches (propagation of forced sound-mode) has been
done in Refs. [21-22]. Here, we only consider the one-dimensional
propagation of plane wave by neglecting the complicated real
boundary conditions.
\newline We assume that the system of hard-disk gas (star) is composed of identical
particles (stars) of the same mass. The velocities of these
particles are restricted to, e.g., : ${\bf v}_1, {\bf v}_2, \cdots,
{\bf v}_p$, $p$ is a finite positive integer. The discrete number
density of particles are denoted by $N_i ({\bf x},t)$ associated
with the velocity ${\bf v}_i$ at point ${\bf x}$ and time $t$.
\newline This simplified model, i.e., the $2\times n$-velocity
model, is to consider a one-component discrete velocity (rotating)
disk- (or vortex-)gas such that the particles can attain 2$n$
velocities in the 2D-plane. In particular, the velocity
discretization is characterized by
\newline (i) $|{\bf v}_i |= c$,
\newline (ii) ${\bf v}_i +{\bf v}_{i+n}= 0$, \newline (iii) ${\bf
v}_i \cdot {\bf v}_{i+1}$= $c^2 \cos (\pi/n)$, $i=1,\cdots, 2n$;
\newline where the index is to be intended modulo 2$n$, i.e.
$i\equiv i+2n$. Such a model is called the planar 2$n$-velocity
model. If only elastic collisions are taken into account, then the
non-trivial admissible ones (where this term is used to denote those
collisions which produce non-vanishing terms in the collision
operator) are [21-22]
\begin{displaymath} 
({\bf v}_i,{\bf v}_{i+n})
  \longleftrightarrow ({\bf v}_j,{\bf v}_{j+n}) \hspace*{6mm} \forall j \not= i,
  i=1,\cdots, 2n.
\end{displaymath}
Besides, the momentum and energy are presumably preserved
\begin{displaymath}
 {\bf v}_i+{\bf v}_{i+n}={\bf v}_j+{\bf v}_{j+n} ,
\end{displaymath}
\begin{displaymath}
 {\mid {\bf v}_i \mid}^2+{\mid {\bf v}_{i+n} \mid}^2={\mid {\bf v}_j \mid}^2
  +{\mid {\bf v}_{j+n} \mid}^2 .
\end{displaymath}
Moreover, all the velocity directions after collisions are assumed
to be equally probable. We have no intentions to consider those
unstable encounter/departure for two-(rotating) vortex (or disk)
collisions [17-18,29] (please see also the references
therein).\newline
Considering binary (two-disk encounter each time) collision only,
the equation of discrete kinetic models proposed in Refs. [21-22,28]
is a system of $2n(=p)$ semilinear partial differential equations of
the hyperbolic type :
\begin{equation}
 \frac{\partial}{\partial t}N_i +{\bf v}_i \cdot\frac{\partial}{\partial
 {\bf x}} N_i =\frac{2 c S}{n} \sum_{j=1}^{n} N_j N_{j+n}-N_i
 N_{i+n} , \hspace*{6mm} i=1,\cdots, 2 n,
\end{equation}
where $N_i=N_{i+2n}$ are unknown functions, and ${\bf v}_i$ =$ c
(\cos[\theta+(i-1) \pi/n], \sin[\theta+(i-1)\pi/n])$; $c$ is a
reference velocity modulus, $S$ is an effective collision
cross-section for the 2-(rotating)disk system [21-22,28], $\theta$
is the free orientation parameter (the orientation starting from the
positive $x$-axis to the $v_1$ direction and is relevant to the
(net) induced scattering measured relative to the sound-propagating
direction) which might be linked to the external field or the
angular momentum or the rotation effects.
\newline Since passage of the sound wave causes a small departure
from equilibrium (Maxwellian type) resulting in energy loss owing to
internal friction and heat conduction, we linearize above equations
around a uniform Maxwellian state ($N_0$) by setting $N_i (t,x)$
=$N_0$ $(1+P_i (t,x))$, where $P_i$ is a small perturbation. The
Maxwellian here is presumed to be the same as in Refs. [20-21].
After some manipulations we then have
\begin{equation}
 [\frac{\partial^2 }{\partial t^2} +c^2 \cos^2 [\theta+\frac{(m-1)\pi}{n}]
 \frac{\partial^2 }{\partial x^2} +4 c S N_0 \frac{\partial
 }{\partial t}] D_m= \frac{4 c S N_0}{n} \sum_{k=1}^{n} \frac{\partial
 }{\partial t} D_k  ,
\end{equation}
where $D_m =(P_m +P_{m+n})/2$, $m=1,\cdots,n$, since $D_1 =D_m$ for
$1=m$ (mod $2 n)$.
We are ready to look for the solutions in the form of plane wave
$D_m$= $a_m$ exp $i (k x- \omega t)$, $(m=1,\cdots,n)$, with
$\omega$=$\omega(k)$. This is related to the dispersion relations
of 1D forced ultrasound propagation of dilute monatomic
hard-sphere gases problem. So we have
\begin{equation}
 \{1+i h-2 \lambda^2 \cos^2 [\theta+\frac{(m-1)\pi}{n}]\} a_m -\frac{i h}{n}
 \sum_{k=1}^n a_k =0  , \hspace*{10mm} m=1,\cdots,n,
\end{equation}
where $\lambda=k c/(\sqrt{2}\omega)$, $h=4 c S N_0/\omega\propto$
1/K$_n$ is the rarefaction parameter of the gas; K$_n$ is the Knudsen
number which is defined as the ratio of the mean free path of hard-disk
or vortex
gases to the wave length of the plane sound wave . \newline Let
$a_m$ = ${\cal{C}}/(1+i h-2 \lambda^2 \cos^2 [\theta+(m-1)\pi/n])$,
where ${\cal{C}}$ is an arbitrary, unknown constant, since we here
only have interest in the eigenvalues of above relation. The
eigenvalue problems for different $2\times n$-velocity model reduces
to $F_n $ $(\lambda) =0$, or
\begin{equation}
  1-\frac{i h}{n} \sum^n_{m=1} \frac{1}{1+i h-2 \lambda^2
  \cos^2\,[\theta+\frac{(m-1)\pi}{n}]} =0.
\end{equation}
We solve $n=2$ case, i.e., 4-velocity case. The admissible collision
: $(1,3)\longleftrightarrow (2,4)$ for system of rotating disks or
vortex gases during binary encounter is shown schematically in Fig.
3. The corresponding eigenvalue equations become algebraic
polynomial-form with the complex roots being the results of
$\lambda$s. \newline For 2$\times$2-velocity model, we obtain
$\hspace*{3mm}
 1-(i h/2) \sum^2_{m=1} \{1/(1+i h-2 \lambda^2  \cos^2\,[\theta+(m-1)\pi/2 ])\} =0
 $. \newline 
\section{Numerical Results and Discussions}
By using the standard mathematical or numerical software, e.g.
Mathematica or Matlab, we can obtain the complex roots
($\lambda=\lambda_r +$ i $\lambda_i$) for the polynomial equations
above [21-22]. The roots are the values for the nondimensionalized
dispersion (positive real part) and the attenuation or absorption
(positive imaginary part), respectively. We plot those of $\theta=0$
into Fig. 4. Curves of branch I follows the conventional dispersion
relation of ultrasound propagation in dilute monatomic hard-sphere
gases [21-22]. Our results could also be applied to the case : as
one rotating disk (star or vortex) colliding with the plane boundary
(the image disk (star or vortex) moving in the opposite direction
within the boundary domain and approaching to the boundary too
[17-18,29]) after then, i.e. the at-a-distance encounter with the
impact parameter being small but finite, they depart (a grazing
collision [30]) following the routes perpendicular to their original
direction.
\newline From a modern point of view, dissipations of the sound wave
arise fundamentally because of a necessary coupling between density
and energy fluctuations induced by disturbances. Within one mean
free path or so of an oscillating boundary, a free molecular flow
solution can probably be computed. The damping will quite likely
turn out to be linear because the damping mechanism is the shift in
phase of particles which hit the wall at different times. It seems
there is a classical (dynamical) wave-localization around $h \sim 1$
(cf. (Chu, 2001) in [21]). To conclude for the results of sound
mode, it was observed that, whereas the continuum-mechanic approach
provides a good modeling at low frequencies, it is definitely not
adequate at high frequencies $h \le 2$. Especially the zero
dispersion (phase speed) as $h$ approaches zero [21-22]. As the
wavelength of sound is made significantly shorter, so that the
effects of viscosity and the heat conduction are no longer small,
the validity of continuum-mechanic approach itself becomes
questionable. If there is no rarefaction effect ($h=0$), we have
only real roots for all the models [21-22]. Once $h \not=0$, the
imaginary part appears and the spectra diagram for each model looks
entirely different. In short, the dispersion ($k_r
c/(\sqrt{2}\omega)$) reaches a continuum-value of $1$ for the
4-velocity models once $h$ increases to infinity. The attenuation
($k_i c/(\sqrt{2}\omega)$) for the same models, instead, firstly
increases up to $h\sim 1.7$, then starts to decrease as $h$
increases furthermore. \newline Curves of branch II, however, show a
entirely different trend. The dispersion part seems to follow the
diffusion mode reported in Refs. [21,31]. It increases but never
reaches to a limit. The anomalous attenuation might be due to, if
any, the intrinsic resonance (an {\it eigen-oscillation}) [32] or
the implicit behavior of angular momentum relation during
2-(rotating)disk (or vortex) encounter (each with opposite-sign
rotating direction or vorticity or angular momenta so that the total
(net) vorticity or angular momenta for this two-body encounter is
zero) since the latter is absent or of no need in the formulation of
2-body collisions. We note that the vorticity or direction of the
rotational axis of each disk or vortex (ready to encounter) for the
collision of 2-(rotating)disk (or vortex) system might be in
opposite sign instead of the same sign. We don't know yet at present
whether the former or the latter can favor the anomalous attenuation
? However, the cosmic ray passing through this acoustic-amplified
channel might be accelerated (neutrinos included)! \newline Some
researchers argued that this anomalous phenomenon (negative
attenuation or amplification) might be linked to the direct or
inverse vortex cascades (in two-dimensional turbulence). Note that,
another possible explanation is due to the presumption of the {\it
microreversibility} in the formulation of equation (1) [21-22,28].
This assumption is valid for the reverse collision in the phase
space and is common in the formulation of the Boltzmann approach
[21-22,28,30]. There is a newest (numerical) report [33] about the
physical reasons for amplification of sound waves which are
discussed in terms of redistribution of acoustic energy and its
potential and kinetic components. The role of acoustic energy
exchange with the mean flow was investigated therein.
Compressibility of the background mean flow is taken into account
and its effect on amplification of acoustic pressure was discussed.
All these just mentioned above could also be valid to the system of
rotating (hard-disk-like) stars!
\newline The results presented here also show the intrinsic
thermodynamic properties of the presumed equilibrium states
corresponding to the collision of rotating disks or vortex gases
during binary encounter [34-35]. The rigorous proof of their
existence, however, will be our future work. \newline To check the
free-orientation effects, we plot those results in Fig. 5. We can
observe, the smaller (absolute values of $\lambda$) branch
(propagation of sound mode) or lower values of $\lambda_i$ in this
figure  which shows a continuous trend as $\theta$ increases toward
$\pi/4$. The acoustical attenuation or absorption  keeps decreasing
as $\theta$ increases from 0. At $\theta = \pi/4$, there is no
attenuation, i.e.  $\lambda_i = 0$. Based on this final
orientation-free results, we could make remarks about those
presented in [2] : the angular distribution of the emitted sound is
fundamentally aspherical once there are rotation, accretion and/or
magnetic fields existing during their propagation (of plane sound
waves). The diffusion modes observed in Fig. 4 or Refs. [21,31]
might also be crucial to the sound pulses radiated from the core
which steepen into shock waves that merge as they propagate into the
outer mantle and deposit their energy and momentum with high
efficiency [2].
 We shall
investigate other relevant problems [36] and role of disk-searching
(linked to massive star formation) [37] in the future.

\newpage
\psfig{file=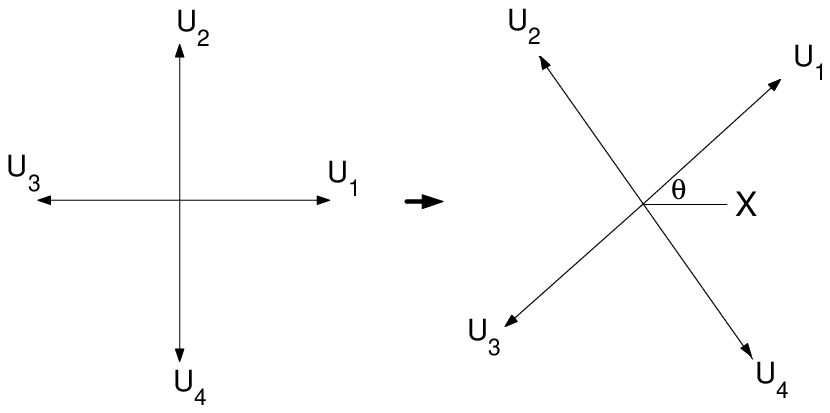,bbllx=0.1cm,bblly=17.5cm,bburx=12cm,bbury=23.8cm,rheight=8cm,rwidth=8cm,clip=}
%
\begin{figure}[h]
\hspace*{6mm} Fig. 3 \hspace*{1mm} Schematic plot for the regular
scattering and the orientational scattering. \newline
\hspace*{7mm} Plane waves propagate along the $X$-direction.
Binary encounters of $U_1$ and $U_3$ and their \newline
\hspace*{7mm} departures  after head-on collisions ($U_2$ and
$U_4$). Number densities $N_i$ are associated to $U_i$.
\end{figure}

\newpage

\psfig{file=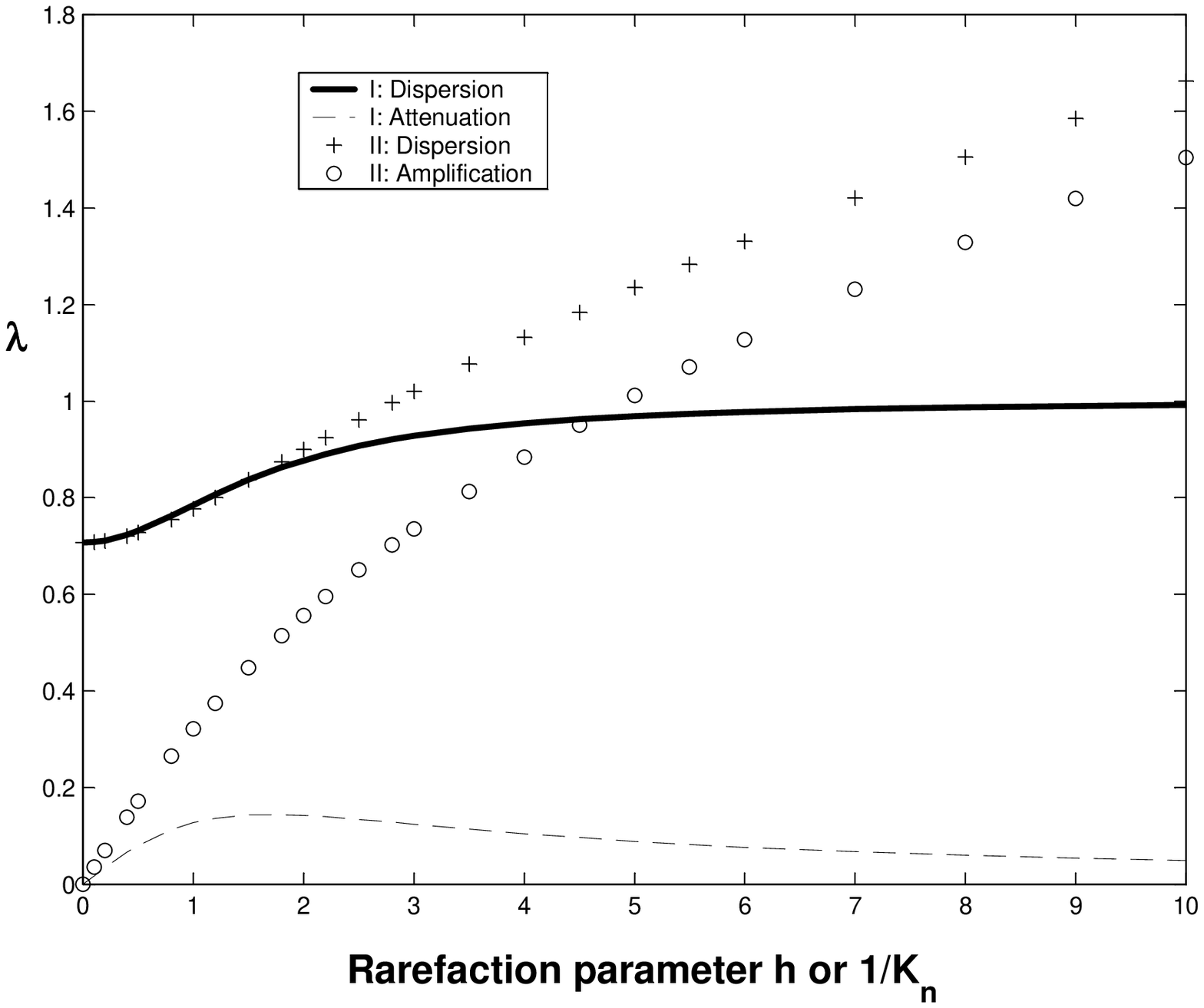,bbllx=2.2cm,bblly=5cm,bburx=20cm,bbury=25cm,rheight=17cm,rwidth=17cm,clip=}
\vspace{10mm}
\begin{figure}[h]
\vspace{6mm}

\hspace*{10mm} Fig. 4 \hspace*{1mm} Dispersion relations of branches
I \& II over a range of $h$ (rarefaction measure). \hspace*{12mm}
$\lambda_r$ : phase speed dispersion, $\lambda_i$ : attenuation or
amplification. ${\bf \lambda}=\lambda_r+$ i $\lambda_i$. $\theta=0$
here.
\end{figure}

\newpage

\psfig{file=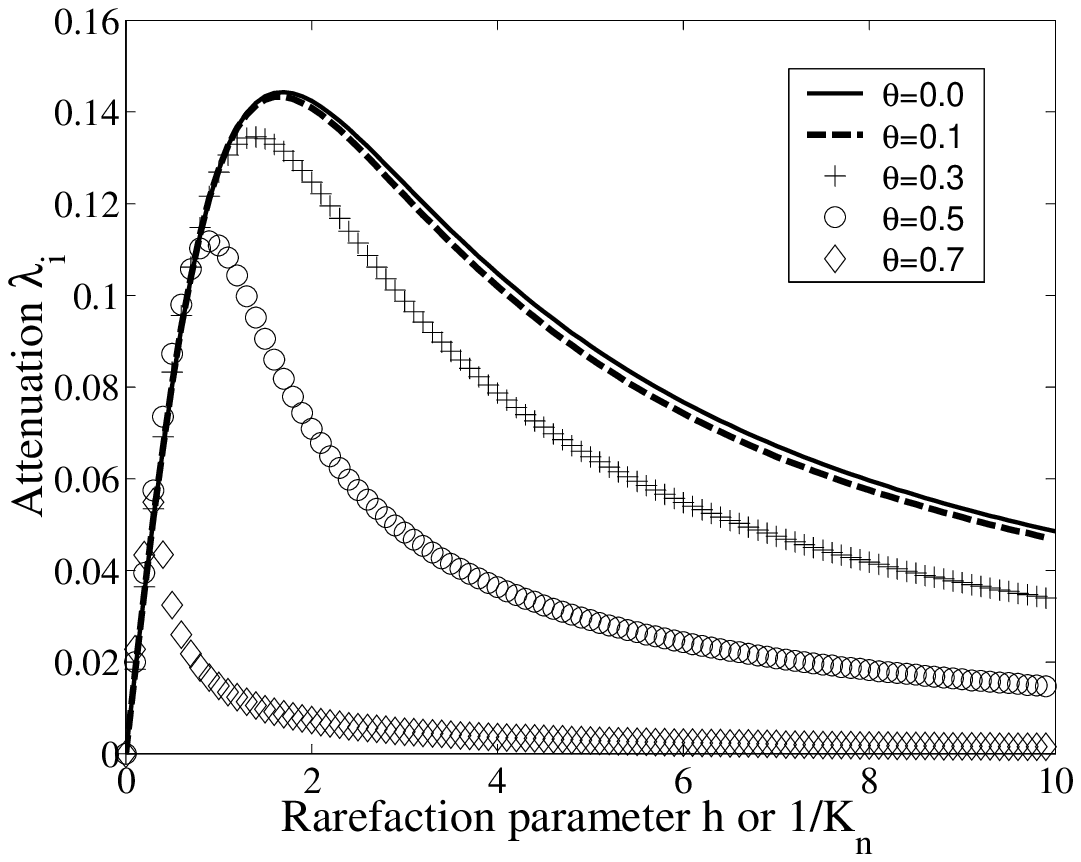,bbllx=0.1cm,bblly=14cm,bburx=12cm,bbury=25cm,rheight=9.6cm,rwidth=9.6cm,clip=}
\begin{figure}[h]
\vspace{6mm}

\hspace*{10mm} Fig. 5 \hspace*{1mm} Free-Orientation effects on the
attenuation or absorption of acoustic \newline \hspace*{7mm} waves
($\lambda_i$). This free orientation might be due to the external
field or rotational effect \newline \hspace*{7mm} (relevant to the
(net) angular momenta existing during the 2-(rotating)disk
encounter).
\end{figure}

\end{document}